\documentclass[11pt,a4paper]{article}

\usepackage{amssymb,dsfont}
\usepackage{slashed}
\usepackage{chet}
\usepackage{tikz}

\newcommand{\be}{\begin{equation}}
\newcommand{\ee}{\end{equation}}
\newcommand{\nn}{\nonumber}

\newcommand{\D}{{\cal D}}
\newcommand{\I}{{\cal I}}
\newcommand{\E}{{\cal E}}

\renewcommand{\O}{{\cal O}}

\newcommand{\rmd}{{\rm d}}
\newcommand{\ts}{\textstyle}

\newcommand{\half}{{\tfrac{1}{2}}}

\newcommand{\pr}{\partial}

\newcommand{\vep}{\varepsilon}
\newcommand{\vphi}{\varphi}

\newcommand{\tr}{{\rm tr}}

\newcommand{\tikzoverset}[2]{%
  \tikz[baseline=(X.base),inner sep=0pt,outer sep=0pt]{%
    \node[inner sep=0pt,outer sep=0pt] (X) {$#2$};
    \node[yshift=2pt] at (X.north) {$#1$};
}}

\preprint{DAMTP 2016/25}

\title{$C_T$ for Non-unitary CFTs in Higher Dimensions}

\author{Hugh Osborn$^{a}$ and Andreas Stergiou$^{b}$
\emails{\href{mailto:ho@damtp.cam.ac.uk}{ho@damtp.cam.ac.uk},
\href{mailto:andreas.stergiou@yale.edu}{andreas.stergiou@yale.edu}}}

\affiliation{$^a$Department of Applied Mathematics and Theoretical
Physics, Wilberforce Road,\\ Cambridge CB3 0WA, England\\
\vspace{3pt}
$^b$Department of Physics, Yale University, New Haven, CT 06520,
USA}

\abstract{The coefficient $C_T$ of the conformal energy-momentum tensor
two-point function is determined for the non-unitary scalar CFTs with four-
and six-derivative kinetic terms. The results match those expected from
large-$N$ calculations for the CFTs arising from the $O(N)$ non-linear
sigma and Gross--Neveu models in specific even dimensions. $C_T$ is also
calculated for the CFT arising from $(n-1)$-form gauge fields with
derivatives in $2n+2$ dimensions. Results for $(n-1)$-form theory extended
to general dimensions as a non-gauge-invariant CFT are also obtained; the
resulting $C_T$ differs from that for the gauge-invariant theory. The
construction of conformal primaries by subtracting descendants of
lower-dimension primaries is also discussed. For free theories this also
leads to an alternative construction of the energy-momentum tensor, which
can be quite involved for higher-derivative theories.}

\date{March 2016}

\begin{document}

\maketitle

\section{Introduction}

It is a truth almost universally acknowledged that there are no non-trivial
unitary conformal field theories in more than six dimensions. Indeed for
superconformal theories this is a long established result \cite{Nahm}, and
it is often conjectured that the superconformal $(2,0)$ theory in six
dimensions is the one theory to rule them all and in the light bind them.

However, there is now good evidence that there are interacting conformal
theories, which contain an energy-momentum tensor, in more than four
dimensions and, indeed, for non-unitary CFTs, for dimensions larger than
six.  The $O(N)$ non-linear sigma model has a tractable $1/N$ expansion
without restriction on the spatial dimension $d$ \cite{Vasiliev1,
Vasiliev2, Vasiliev3}, defining a conformal theory with calculable scaling
dimensions, at least for sufficiently large $N$ \cite{Kleb1}. This suggests
a non-trivial five-dimensional CFT which is also accessible by
$\vep$-expansion methods starting from $4+\vep$ and $6-\vep$ dimensions.
Generalisations to higher dimensions were recently explored
in~\cite{Stergiou6, Gracey}.  To leading order in $1/N$ the non-linear
sigma model comprises an $N$-component scalar $\vphi_i$ with dimension
$\half(d-2)$ and also a singlet $\sigma$ with dimension $2$, clearly
violating the unitarity bound for scalars when $d>6$.

Apart from the scaling dimensions for conformal primary operators and the
parameters determining the three-point functions and, hence, the operator
product expansion, crucial data defining a CFT are given by the correlation
functions involving the energy-momentum tensor.  In any CFT $C_T$, the
coefficient of the two-point function which is fixed up to an overall
constant by conformal invariance, plays a crucial role.  The scale of the
energy-momentum tensor is determined by Ward identities and with our
conventions
\eqn{
S_d{\hspace{-1pt}}^2  \big \langle T^{\mu\nu}(x) \, T^{\sigma\rho}(0)
\big \rangle =
C_T \; \frac{1}{(x^2)^d} \; \I^{\mu\nu,\sigma\rho}    (x)  \, ,
\label{Tdef}}[]
for $S_d=2 \pi^{\frac12 d}/\Gamma(\frac12 d)$ and where $\I$ is the
inversion tensor for symmetric traceless tensors, constructed in terms of
the inversion tensor $I$ for vectors
\eqn{
 \I^{\mu\nu,\sigma\rho} = \half ( I^{\mu\sigma} I^{\nu \rho}
 + I^{\mu\rho} I^{\nu \sigma} ) - \frac{1}{d} \, \eta^{\mu\nu}
 \eta^{\sigma\rho} \, , \quad
I^{\mu\nu}(x) = \eta^{\mu \nu} - \frac{2}{x^2} \, x^\mu x^\nu \,.
\label{Iinv}}[]

$C_T$ may be regarded as a measure of the number of degrees of freedom. It
determines the contribution of the energy-momentum tensor in the conformal
partial-wave expansion, and so is readily determined in bootstrap
calculations.  For the conventional free scalar and fermion theories $C_T$
was determined for arbitrary $d$ and for vector gauge theories for $d=4$
some time ago in \cite{Pet}, and later $C_T$ was calculated for
$(n-1)$-form gauge in $d=2n$ dimensions in \cite{Buchel}. For the $O(N)$
sigma model results for $C_T$ to first order in $1/N$ were obtained by
Petkou \cite{Petkou1, Petkou2} by applying the operator product expansion
to the four-point function for $\phi_i$, and have recently been rederived
by direct calculation and extended to the Gross--Neveu model in
\cite{Kleb4}.  In the non-linear sigma model
\be
C^{O(N)}_{T} = C_{T,S} \Big ( N + C^{O(N)}_{T,1} + {\rm O}(N^{-1}) \Big )
\, ,\qquad C_{T,S} = \frac{d}{d-1} \, ,
\label{Cfree}
\ee
where $C_{T,S}$ is the result for a free scalar in $d$ dimensions. For
general $d$, $C^{O(N)}_{T,1}$ depends on the digamma function,
$\psi(x)=\Gamma'(x)/\Gamma(x)$ but for $d=4+2p$ only the contribution of
the poles $\psi(x) \sim - 1/(x+n), \ n=0,1,\dots$ for some $n$ are
relevant. In consequence, the result for general $d$ reduces to
\be
C^{O(N)}_{T,1}\Big |_{d=4+2p} = (-1)^{p-1} \, \frac{4\, (2p+1)!}
{(p-1)! \, (p+3)!} \, ,
\label{C4p}
\ee
which is an integer~\cite{Stergiou6}. Thus, $C^{O(N)}_{T,1} \big |_{d=4} =0
$, since then the theory reduces to $N$ free scalars with $\sigma$
non-dynamical, whereas  $C^{O(N)}_{T,1} \big |_{d=6} =1 $. The extra $1$
was interpreted in \cite{Kleb1} as the contribution of the dynamical free
scalar $\sigma$ and for $d=6-\vep$ the $\vep$-expansion defines a CFT at a
fixed point starting from the renormalisable $O(N)$-invariant Lagrangian
\be
{\cal L}_6 = -\half \big ( \pr^\mu \vphi_i \pr_\mu \vphi_i
+ \pr^\mu \sigma \pr_\mu \sigma +
g \, \sigma \, \vphi_i \vphi_i \big ) - \tfrac16 \lambda \, \sigma^3 \, ,
\label{Lsix}
\ee
with $g,\lambda = {\rm O}(\vep)$. In higher even dimensions there are
corresponding renormalisable Lagrangians with higher-derivative kinetic
terms for $\sigma$.  For $d=8-\vep$ there is a perturbative fixed point
starting from
\be
{\cal L}_8 = - \tfrac12 \big ( \partial^\mu\vphi_i \partial_\mu \vphi_i
+ \partial^2 \sigma
\partial^2 \sigma + g \, \sigma \vphi_i \vphi_i + \lambda' \, \sigma^2
\partial^2 \sigma \big )
- \tfrac{1}{24} \lambda\, \sigma^4 \, .
\label{Leight}
\ee
The $\beta$ functions for the couplings $g,\lambda',\lambda$ have recently
been calculated by Gracey \cite{Gracey}. In this case $C^{O(N)}_{T,1} \big
|_{d=8} =-4$, which we show arises from the higher-derivative $\sigma$
contribution.

A similar narrative emerges for the Gross--Neveu model with $N$ fermion
fields $\psi_i$. There is also a self-consistent $1/N$ expansion as a
conformal field theory for any dimension $d$. To leading order $\psi_i$ has
dimension $\half(d-1)$, and there is a singlet scalar field $\sigma$ with
scale dimension $1$, which is consequently below the unitary bound for
$d>4$. The leading $1/N$ correction to $C_T$ has been recently determined
in \cite{Kleb4}, and can be expressed in the form
\be
C^{\rm GN}_T = \half d \,  \tr( {\mathds{1}} ) \, N +
C_{T,S} \big ( C^{\rm GN}_{T,1} + {\rm O}(N^{-1}) \big ) \, ,
\ee
where $\half d \, \tr( {\mathds{1}} ) $ is the contribution to $C_T$ for a
single fermion with $\tr( {\mathds{1}} ) $ the sum over spinorial indices.
For general $d$ the result \cite{Kleb4} for  $C^{\rm GN}_{T,1} $ is similar
to that for the sigma model; for even dimensions it reduces to
\be C^{{\rm
GN}}_{T,1}\Big |_{d=2+2p} = (-1)^{p-1} \, \frac{(2p+1)!}{(p-1)! \, (p+2)!}
\, .
\label{C2p}
\ee
In this case $C^{{\rm GN}}_{T,1}\big |_{d=4} = 1$, representing the
contribution of the dynamical scalar $\sigma$ whereas $C^{{\rm
GN}}_{T,1}\big |_{d=6} = -5$. For $d=4-\vep$ equivalent results can be
obtained as a perturbative  $\vep$ expansion at the RG fixed point starting
from the renormalisable Lagrangian
\be
{\cal L}_{{\rm GN},4} = - {\bar \psi}\, {\slashed \pr}   \psi -
\half \, \pr^\mu \sigma \pr_\mu \sigma -
g \, \sigma \, {\bar \psi} \psi - \tfrac{1}{24}  \lambda \, \sigma^4 \, ,
\label{Lfour}
\ee
with $N$ Dirac fields $\psi$.

In this paper we calculate the contributions to $C_T$ corresponding to
higher-derivative scalars, such as that for $\sigma$ in \eqref{Leight}, for
general $d$. The energy-momentum tensor is determined from the
corresponding local Weyl-invariant actions on curved space quadratic in a
scalar field $\vphi$. The construction of such actions is equivalent to
obtaining conformal differential operators starting from powers of the
Laplacian. We then determine $C_{T}$ for scalar theories with kinetic terms
with $2p$ derivatives for $p=2,3$, and conjecture a result for general
$p$.\foot{$C_{T,S}$ in \eqref{Cfree} corresponds to $p=1$.} The formula
agrees with \eqref{C4p} and \eqref{C2p} for the relevant values of $d$.

In section 3 we consider $(n-1)$-form gauge theories  with additional
derivatives  in $2n+2$ dimensions when they also define a CFT and obtain
$C_T$ in this case. Reflecting the lack of unitarity, $C_T<0$.  We also
discuss in section 4 the $(n-1)$-form theory, without additional
derivatives, extended to define a CFT away from $d=2n$ dimensions when
gauge invariance is lost.

The energy-momentum tensors in the higher-derivative theories are rather
non-trivial. In section 5 we discuss an alternative construction which
constructs a spin-two conformal primary by successively subtracting the
descendants of lower-dimension primary operators. The final expressions
thereby obtained are identical with those derived from curved-space
Weyl-invariant actions; the subtractions are related to improvement terms
which need to be added to the canonical energy-momentum tensor to obtain a
tensor which is traceless as well as symmetric.

\section{Higher-derivative Scalar Theories}

The actions for higher-derivative free scalars considered here have the
form
\eqn{S_{4} [\vphi] = - \int \rmd^d x \; \half \, \pr^2  \vphi \, \pr^2
\vphi \, , \qquad S_6[\vphi] = - \int \rmd^d x \; \half \, \pr^\mu \pr^2
\vphi \, \pr_\mu \pr^2 \vphi \, .  \label{Sphi}}[]
Theories starting from such actions were considered in \cite{David,David2}.
A symmetric traceless energy-momentum tensor may be obtained by the usual
Noether procedure or by extending \eqref{Sphi} to a general curved space
background so as to be invariant under Weyl rescalings of the metric. Assuming
diffeomorphism invariance then reducing to flat space ensures that the resulting
energy momentum tensor satisfies conformal Ward identities which ensure that
it is a conformal primary.

For $S_4$ the extension to a Weyl invariant form on curved space
 is equivalent to constructing the Paneitz operator
\cite{Paneitz} (see also \cite{Fradkin1} and \cite{Riegert} for the $d=4$
version of the Paneitz operator) and for $S_6$ this involves the
generalisation of the $d=6$ Branson operator \cite{Branson} to general $d$.
These operators provide extensions of $\nabla^2 \nabla^2$ and $-\nabla^2
\nabla^2 \nabla^2$ to conformal differential operators. A convenient form
for the Branson operator for general $d$ was constructed in
\cite{6dimP}\footnote{See version 3 of \cite{6dimP}.} by extending $S_6$ to
a Weyl invariant form on an arbitrary curved background.  A useful
mathematical discussion for arbitrary powers of the Laplacian is contained
in  \cite{Gover}, such operators fail to exist in particular integer
dimensions, for the Paneitz, Branson operators these are
$d=2,4$.\footnote{This is then an obstruction to relating Weyl and
conformal invariance \cite{WeylC}, but it also prevents the existence of a
symmetric traceless energy-momentum tensor which is a conformal primary.
Related discussions are given in \cite{Rajabpour} and for $d=2$ in
\cite{Wiese,Nakayama}.} Varying the metric about flat space gives (an
alternative derivation based on a generalised Noether procedure is given in
\cite{CFTNotes})
\eqna{T_{\vphi,4}^{\mu\nu}  =  {}& 2\, \pr^\mu \pr^\nu \!  \vphi \, \pr^2 \vphi - \half \, \eta^{\mu\nu} \,
\pr^2 \vphi \pr^2 \vphi  - \pr^\mu ( \pr^\nu \vphi \, \pr^2 \vphi )
- \pr^\nu ( \pr^\mu \vphi \, \pr^2 \vphi ) + \eta^{\mu\nu} \, \pr_\rho ( \pr^\rho \vphi \, \pr^2 \vphi) \\
 &  {}   +
2\, \D^{\mu\nu\sigma\rho} \big ( \pr_\sigma \vphi\, \pr_\rho \vphi \big )
- \frac{1}{d-1} \, (\pr^\mu\pr^\nu - \eta^{\mu\nu}\pr^2)
\big (  \pr^\rho \vphi \, \pr_\rho \vphi
- \half (d-4) \,  \pr^2 \vphi \, \vphi \big ) \, ,
\label{T4phi}
}[]
for
\eqna{\D^{\mu\nu\sigma\rho} = {}&  \frac{1}{d-2}\big (  \eta^{\mu(\sigma}  \pr^{\rho)} \pr^\nu
 +  \eta^{\nu(\sigma}   \pr^{\rho)} \pr^\mu  -
\eta^{\mu(\sigma}\eta^{\rho)\nu}
\pr^2 - \eta^{\mu\nu} \pr^\sigma\pr^\rho \big ) \\
\noalign{\vskip -4pt}
&{}  - \frac{1}{(d-2)(d-1)} \, \big (\pr^\mu\pr^\nu - \eta^{\mu\nu}\pr^2
\big ) \eta^{\sigma\rho} \, ,
}[]
where $\pr_\mu \D^{\mu\nu\sigma\rho} = 0, \,
\eta_{\mu\nu}\D^{\mu\nu\sigma\rho} = - \pr^\sigma \pr^\rho$, and
\eqna{T_{\vphi,6}^{\mu\nu} = {}& \pr^\mu\pr^2 \vphi \, \pr^\nu \pr^2 \vphi  -  2\, \pr^\mu \pr^\nu \!  \vphi \, \pr^2\pr^2 \vphi
 - \half \, \eta^{\mu\nu} \, \pr^\sigma \pr^2 \vphi \, \pr_\sigma\pr^2 \vphi   \\
&{} +  \pr^\mu ( \pr^\nu \vphi \, \pr^2 \pr^2 \vphi )
   + \pr^\nu ( \pr^\mu \vphi \, \pr^2\pr^2  \vphi ) - \eta^{\mu\nu} \, \pr_\rho ( \pr^\rho \vphi \, \pr^2\pr^2  \vphi)  \\
 &\,  {}   + 8 \,  \D^{\mu\nu\sigma\rho} ( \pr_\sigma \pr_\rho \vphi \, \pr^2 \vphi)
- \frac{1}{d-1} \, \big (\pr^\mu\pr^\nu - \eta^{\mu\nu}\pr^2 \big ) O  \\
&{}+ \lambda\, \D_B^{\mu\nu\sigma\rho}\, (\pr_\sigma \vphi \pr_\rho \vphi) \, ,
 \\
\noalign{\vskip 4pt}
O = {}&  \half (d-6)\, \pr^2  ( \pr^2 \vphi \, \vphi) +
(10-d) \,  \pr_\rho ( \pr^\rho  \vphi  \, \pr^2 \vphi)
+ \tfrac34 (d-2) \, \pr^2 \vphi \, \pr^2 \vphi \, ,
\label{T6phi}}[]
where
\eqna{\D_B^{\mu\nu\sigma\rho} = {}&  \D^{\mu\nu\sigma\rho}\,  \pr^2
- \frac{1}{d-1} \,  \big (\pr^\mu\pr^\nu - \eta^{\mu\nu}\pr^2 \big )
\pr^\sigma \pr^\rho  \, .}[]
It is useful to note that
\eqn{\D_B^{\mu\nu\sigma\rho} (\pr_\sigma v_\rho + \pr_\rho v_\sigma) = 0
\, , \quad
\D_B^{\mu\nu\sigma\rho} \eta_{\sigma\rho} = 0 \, , \quad
\pr_\mu  \D_B^{\mu\nu\sigma\rho} = 0 \, , \quad \eta_{\mu\nu}
\D_B^{\mu\nu\sigma\rho}  = 0 \, .}[]
The terms in the expressions for $T_{\vphi,4}^{\mu\nu}, \,
T_{\vphi,6}^{\mu\nu}$ involving the second- and higher-order derivative
operators $\D^{\mu\nu\sigma\rho}, \, \pr^\mu\pr^\nu - \eta^{\mu\nu}\pr^2 $
and $ \D_B^{\mu\nu\sigma\rho}$ arise from explicit curvature-dependent
terms in the curved-space action and represent improvement terms to be
added to the canonical energy-momentum tensor. In particular the
contribution involving $ \D_B^{\mu\nu\sigma\rho}$ comes from the reduction
of a term in the curved-space result proportional to $\pr_\mu \vphi \pr_\nu
\vphi \, B^{\mu\nu}$, with $B^{\mu\nu}$ the Bach tensor, and this gives
\eqn{\lambda = - \frac{8}{d-4} \, .
\label{defl}}[]

The results for $T_{\vphi,4}^{\mu\nu}$ and
$T_{\vphi,6}^{\mu\nu}$ in \eqref{T4phi} and
\eqref{T6phi} obey the conservation and trace conditions,
\eqn{\pr_\mu T_{\vphi,2p}^{\mu\nu}
= (-1)^{p-1} (\pr^2)^p  \vphi \, \pr^\nu  \vphi \, , \quad
 \eta_{\mu\nu} T_{\vphi,2p}^{\mu\nu}
=   (-1)^{p-1}\Delta_{2p} \, (\pr^2)^p \vphi\,  \vphi \, , \quad
\Delta_{2p}  = \half (d-2 p) \, ,
}[]
which of course vanish on the relevant equations of motion
$ (\pr^2)^p \vphi =0$.

Correlators and operator products in the free field theories are
determined just by
\eqna{\big \langle \vphi(x) \, \vphi(0) \big  \rangle_4 = {}&
\frac{1}{2(d-4)(d-2)S_d} \,
\frac{1}{(x^2)^{\frac12(d-4)} }\,  ,  \\
\big \langle \vphi(x) \, \vphi(0) \big  \rangle_6 = {}&
\frac{1}{8(d-6)(d-4)(d-2)S_d} \,
\frac{1}{(x^2)^{\frac12(d-6)} }\, .}[opeI]
These are respectively
singular when $d\to 4,6$ but in \eqref{T4phi}, \eqref{T6phi} the only terms
not involving $\pr \vphi$ have overall factors $d-4, \, d-6$ in each case.
From this term, for both $T_{\vphi,4}^{\mu\nu}, \,
T_{\vphi,6}^{\mu\nu}$, we may verify
for the leading term in the operator product
\eqn{
S_d \, T_{\vphi,2p}^{\mu\nu} (x) \, \vphi(0) \sim -
\frac{d\, \Delta_{2p} }{d-1} \,
\frac{1}{(x^2)^{\frac12 d}} \, \bigg ( \frac{x^\mu x^\nu}{x^2} -
\frac{1}{d} \, \eta^{\mu\nu} \bigg ) \vphi(0) \, .}[]
The coefficient is determined by Ward identities assuming
$ T_{\vphi,2p}^{\hskip 1pt \mu\nu} $ is canonically normalised.

In free field theories any local operator formed from $\vphi$ with
derivatives at the same point can be decomposed in terms of conformal
primaries and descendants, or derivatives, of conformal primaries of lower
dimension. Since $T^{\mu\nu}$ is a conformal primary the result in \eqref{Tdef} is
therefore unchanged for $T^{\sigma\rho} \to T^{\sigma\rho} + \pr_\tau
X^{\sigma\rho\tau}$ for any local $X^{\sigma\rho\tau}$ expressible as a
conformal primary or descendant.  This ensures that, dropping also terms
which vanish on the equations of motion,
\eqna{\big \langle T_{\vphi,4}^{\hskip 1pt \mu\nu}(x) \,
T_{\vphi,4}^{\hskip 1pt \sigma\rho}(0)  \big \rangle = {}& 2\,
\big \langle T_{\vphi,4}^{\hskip 1pt \mu\nu}(x) \,
\pr^{\sigma}\pr^{\rho}\vphi \pr^2 \vphi (0)\big \rangle \, ,   \\
\big \langle T_{\vphi,6}^{\hskip 1pt \mu\nu}(x) \,
T_{\vphi,6}^{\hskip 1pt \sigma\rho}(0)  \big \rangle = {}&
-3 \, \big \langle T_{\vphi,6}^{\hskip 1pt \mu\nu}(x) \,
\pr^{\sigma}\pr^{\rho}\vphi \pr^2\pr^2 \vphi (0)\big \rangle
= 3 \, \big \langle T_{\vphi,6}^{\hskip 1pt \mu\nu}(x) \,  \pr^{\sigma}
\pr^2 \vphi \pr^\rho \pr^2 \vphi (0)\big \rangle\, .}[]
We get, with $\lambda$ as in \eqref{defl},
\eqn{
C_{T,\vphi,4} = - \frac{2d(d+4)}{(d-2)(d-1)}\, , \qquad
C_{T,\vphi,6} = \frac{3 d (d+4)(d+6)}{(d-4)(d-2)(d-1)}  \, .
\label{resCT}}[]

There is an obvious generalisation of \eqref{Sphi} to actions with more
derivatives, $S_{2p}$ formed from $(\pr^2)^r \vphi$ or $\pr_\mu  (\pr^2)^r
\vphi$ for $p=2r$ or $p=2r+1$.  On the basis of the results in
\eqref{resCT} we may guess
\eqn{C_{T,\vphi,2p} = C_{T,S} \,
\frac{p\, (\half d +2 )_{p-1}}
{(-\half d + 1 )_{p-1}} \,,\qquad p=1,2,\ldots\,,
\label{Cscalar}
}[]
where $(a)_n=\Gamma(a+n)/\Gamma(a)$ is the Pochhammer symbol
and $C_{T,S}$ the free scalar result given in  \eqref{Cfree}.
The result \eqref{Cscalar} agrees with \eqref{C4p}, \eqref{C2p} for
the particular cases of $d$.

\section{Higher-order \texorpdfstring{$(n-1)$}{(n-1)}-Form Gauge Theories}
In $d=2n$ dimensions free conformal theories can be formed from
$(n-1)$-form gauge fields $A_{\mu_1 \dots \mu_{n-1}}$ with $n$-form field
strengths $F_{\mu_1 \ldots \mu_{n} } = n\,  \pr\raisebox{-1.5 pt}
{$\scriptstyle {[\mu_1}$} A\raisebox{-1.5 pt} {$\scriptstyle {\mu_2 \dots
\mu_n]}$}$ with the gauge invariant action
$S_{n,0}[A]
= - \frac{1}{2\,  n!} \,  \int \rmd^{2n+2} x  \;
 F^{\mu_1 \dots \mu_{n}} F_{\mu_1 \dots \mu_{n}}$,
generalising the conformal invariant Maxwell theory in four
dimensions.
Here we consider the corresponding theory with two additional
derivatives given by the action in $d=2n+2$ dimensions
\eqna{S_{n,2}[A] ={}&  - \frac{1}{2n!} \,  \int \rmd^{2n+2} x \; \pr^\lambda
F^{\mu_1 \dots \mu_n} \pr_\lambda F_{\mu_1 \dots \mu_n}   \\ = {}&-
\frac{1}{2(n-1)!} \,  \int \rmd^{2n+2} x \; \pr_\lambda F^{\lambda\mu_1
\dots \mu_{n-1}}  \pr^\rho F_{\rho\mu_1 \dots \mu_{n-1}}
 \, .
\label{SA}}[] This may be extended to a general curved background metric
$\gamma_{\mu\nu}$ so as to be invariant under Weyl rescalings in the form
\eqna{S_{n,2}[A]
= {}&- \frac{1}{2\,  n!} \,  \int \rmd^{2n+2} x \sqrt{-\gamma} \; \Big (
\nabla^\lambda F^{\mu_1 \dots \mu_{n}}  \nabla_\lambda F_{\mu_1 \dots \mu_{n}}
 \\
\noalign{\vskip-6pt}
&\hskip 4.3cm{}
+ \big ( 2n \, P_{\lambda\rho} + (n+2)  \, \gamma_{\lambda\rho} \, {\hat R} \big )
F^{\lambda\mu_1 \dots \mu_{n-1}}   F^{\rho}{}_{\mu_1 \dots \mu_{n-1}}  \\
\noalign{\vskip-1pt}
&\hskip 4.3cm{}
+ c_W \,  n(n-1) \, W_{\mu\nu\lambda\rho} \,
F^{\mu\nu\mu_1 \dots \mu_{n-2}}   F^{\lambda \rho}{}_{\mu_1 \dots \mu_{n-2 }}\Big )
 \, ,
\label{SA2}}[]
where $P_{\lambda\rho} = \frac{1}{d-2} ( R_{\lambda\rho} -
\gamma_{\lambda\rho}\,  {\hat R} )$ is the Schouten tensor, ${\hat R} =
\frac{1}{2(d-1)} R$ a rescaled scalar curvature and $W_{\mu\nu\lambda\rho}$
the Weyl tensor. The Weyl tensor term is invariant under Weyl rescaling by
itself and so, for $n>1$, has an arbitrary coefficient. The expression for
the action may be written in various forms with the aid of the identity,
for $d$ arbitrary,
\eqna{n \, \nabla_\lambda  & \nabla_\rho
\big (F^{\lambda\mu_1 \dots \mu_{n-1}}   F^{\rho}{}_{\mu_1 \dots \mu_{n-1}} \big )
- \nabla^2 \big (  F^{\mu_1 \dots \mu_{n}}   F_{\mu_1 \dots \mu_{n}} \big )  \\
= {} &  -  \nabla^\lambda F^{\mu_1 \dots \mu_n} \nabla_\lambda F_{\mu_1 \dots \mu_n}
+n\, \nabla_\lambda F^{\lambda\mu_1 \dots \mu_{n-1}}  \nabla^\rho F_{\rho\mu_1 \dots \mu_{n-1}} \\
&{} - n \, \big ( (d-2n)  \, P_{\lambda\rho} +  \gamma_{\lambda\rho} \, {\hat R} \big )
F^{\lambda\mu_1 \dots \mu_{n-1}}   F^{\rho}{}_{\mu_1 \dots \mu_{n-1}}  \\ &{}
+ \half n(n-1) \, W_{\mu\nu\lambda\rho} \,
F^{\mu\nu\mu_1 \dots \mu_{n-2}}   F^{\lambda \rho}{}_{\mu_1 \dots \mu_{n-2 }} \, ,
\label{ncurve}}[]
depending on the Bianchi identity for $F$.\footnote{In $d$-dimensions
there is a conformal scalar
\eqna{(4n+2-d) \big ( (d-2n) \,
\nabla^\lambda F^{\mu_1 \dots \mu_n} \nabla_\lambda F_{\mu_1 \dots \mu_n}
- n \,
\nabla_\lambda F^{\lambda\mu_1 \dots \mu_{n-1}}  \nabla^\rho F_{\rho\mu_1 \dots \mu_{n-1}}
\big )  \\
{} - (n+1)(d-2n) \, ( \nabla^2 - 2n \, {\hat R} )
\, \big ( F^{\mu_1 \dots \mu_n} F_{\mu_1 \dots \mu_n} \big ) \, ,}
which generalises an expression obtained by Parker and Rosenberg
\cite{Parker}.  For $d=4n+2$ this is just the conformal Laplacian acting on
$F^2$.}

Varying the metric in \eqref{ncurve} determines the corresponding flat
space energy momentum tensor for $(n-1)$-form gauge fields involving two
derivatives
\eqna{n! \,T_{n,2}^{\mu\nu} ={}& n\,
\pr^\lambda F^{\mu{\hskip 0.5pt} \mu_1 \dots \mu_{n-1}} \,
\pr_\lambda F^\nu{\!}_{\mu_1 \dots \mu_{n-1}}
+  \pr^\mu F^{\mu_1 \dots \mu_n}\,  \pr^\nu F_{\mu_1 \dots \mu_n} \\
&{} - \half \, \eta^{\mu\nu} \,
 \pr^\lambda F^{\mu_1 \dots \mu_n} \pr_\lambda F_{\mu_1 \dots \mu_n}
  + 2n \, \pr_\lambda \big ( F^{\lambda  \mu_1 \dots \mu_{n-1}}
  {\tikzoverset{\text{\tiny$\leftrightarrow$}}{\pr}}{}^{(\mu}
  F^{\nu)}{}_{\mu_1 \dots \mu_{n-1}}\big )  \\
 &{}
- \half n \, \pr^2
\big (  F^{\mu {\hskip 0.5pt} \mu_1 \dots \mu_{n-1}} F^\nu{}_{\mu_1 \dots \mu_{n-1}}\big )
+ n\, \D^{\mu\nu\sigma\rho} \big ( F_\sigma{\hskip -0.5pt}{}^{\mu_1 \dots \mu_{n-1}}
 F_{\rho{\hskip 0.5pt} \mu_1 \dots \mu_{n-1}}\big )  \\
 &{}- \tfrac{n+2}{2(2n+1)} \, (\pr^\mu \pr^\nu - \eta^{\mu\nu} \pr^2 )
 \big ( F^{\mu_1 \dots \mu_n}  F_{\mu_1 \dots \mu_n} \big )  \\
&{}+ c_W \, n(n-1) \, \E_W{}^{\mu\sigma\nu\rho,\epsilon\eta\kappa\lambda} \,
\pr_\sigma \pr_\rho \big ( F_{\epsilon\eta}{}^{\mu_1 \dots \mu_{n-2}}
F_{\kappa\lambda \hskip 0.5pt \mu_1 \dots \mu_{n-2 }}  \big ) \, ,
\label{EMA}}[]
where ${\tikzoverset{\text{\tiny$\leftrightarrow$}}{\pr}} = \half ( \pr -
{\tikzoverset{\text{\tiny$\leftarrow$}}{\pr}})$.  In the last line $\E_W$  is
the projector for traceless tensors satisfying the symmetries of the Weyl
tensor and has the properties
$\E_W{}^{\mu\sigma\nu\rho,\epsilon\eta\kappa\lambda} =
\E_W{}^{[\mu\sigma][\nu\rho],[\epsilon\eta][\kappa\lambda]} =
\E_W{}^{\epsilon\eta\kappa\lambda,\mu\sigma\nu\rho}$, $ \eta_{\mu\nu} \,
\E_W{}^{\mu\sigma\nu\rho,\epsilon\eta\kappa\lambda} =
\E_W{}^{\mu[\sigma\nu\rho],\epsilon\eta\kappa\lambda} = 0$.  The
energy-momentum tensor in \eqref{EMA}  satisfies, using the Bianchi
identity,
\eqn{n! \, \pr_\mu T_{n,2}^{\mu\nu} = - n \,
\pr^2 \pr_\mu F^{\mu{\hskip 0.5pt} \mu_1 \dots \mu_{n-1}} \,
F^\nu{\!}_{\mu_1 \dots \mu_{n-1}}\, ,
\qquad n! \, \eta_{\mu\nu} T_{n,2}^{\mu\nu} = 0\, ,}[]
and so $T_{n,2}^{\mu\nu}$ is conserved subject to the equation of motion
$\pr^2 \pr_\mu F^{\mu{\hskip 0.5pt} \mu_1 \dots \mu_{n-1}} =0$.

In a Feynman type gauge the action \eqref{SA} reduces to
\eqn{
S_{n,2}[A] =  -  \frac{1}{2(n-1)!}\int \rmd^{2n+2} x \;
\pr^2 A ^{\mu_1
\dots \mu_{n-1}}  \,
\pr^2  A_{\mu_1 \dots \mu_{n-1}} \, ,
\label{SA3}}[]
so that in this gauge
\eqn{\big \langle A_{\mu_1 \dots \mu_{n-1}} (x) \,
 A_{\nu_1 \dots \nu_{n-1}} (0) \big \rangle =
\frac{(n-2)!}{8n\, S_{2n+2}} \, \frac{1}{(x^2)^{n-1}} \,
\E^{(n-1)}{\!}_{\mu_1 \dots \mu_{n-1},\, \nu_1 \dots \nu_{n-1}} \, ,}[]
where
\be
\E^{(n)}{\!}_{\mu_1 \dots \mu_{n},}{}^{\nu_1 \dots \nu_{n}}
= \delta_{[\mu_1}{\!}^{\nu_1} \dots
\delta_{\mu_{n}]}{\!}^{\nu_{n}} \, ,
\ee
is the projector on to rank $n$ antisymmetric tensors.
Then, with $F$ defined in \eqref{SA}, the two-point function for $F$, which
is gauge independent, is given by
\eqn{\big \langle F_{\mu_1 \dots \mu_{n}} (x) \,
F_{\nu_1 \dots \nu_{n}} (0) \big \rangle =
\frac{ n!}{4\, S_{2n+2}} \, \frac{1}{(x^2)^{n}} \,
\E^{(n)}{\!}_{\mu_1 \dots \mu_{n},}{}^{\lambda_1 \dots \lambda_{n}}
I_{\lambda_1\, \nu_1}(x) \dots
I_{\lambda_n \, \nu_{n}}(x)  \, ,
\label{Ftwo}}[]
with $I_{\lambda\hskip0.5pt\nu}(x)$ determined by \eqref{Iinv}.
The result \eqref{Ftwo}
has the expected form for $F_{\mu_1\dots \mu_n}$ a conformal primary of
dimension $n$.

For this theory the two-point function of the energy-momentum tensor
is determined from
\begin{align}
\big \langle T_{n,2}^{\hskip 1pt \mu\nu}(x) \,
T_{n,2}^{\hskip 1pt \sigma\rho}(0)  \big \rangle = {}&
\big \langle T_{n,2}^{\hskip 1pt \mu\nu}(x) \, X_n^{\sigma\rho}(0)
\big \rangle \, ,
\label{TAX}
\end{align}
for, discarding total derivatives and terms which vanish on the equations
of motion, \be n! \,X_n^{\sigma\rho} = n\, \pr^\lambda F^{\sigma{\hskip
0.5pt} \mu_1 \dots \mu_{n-1}} \, \pr_\lambda F^\rho{\!}_{\mu_1 \dots
\mu_{n-1}} +  \pr^\sigma F^{\mu_1 \dots \mu_n}\,  \pr^\rho F_{\mu_1 \dots
\mu_n} \, .  \ee The $c_W$ contribution in \eqref{EMA} can also be dropped
since this term is a conformal primary descendant of a conformal primary
and does not contribute to \eqref{TAX}.  Using the two-point function
\eqref{Ftwo} the combinatorics for arbitrary $n$ can be handled with the
identities
\begin{align}
&{} \E^{(n)}{\!}_{\mu_1 \dots \mu_{n},\,\sigma_1 \dots \sigma_{p}
\lambda_1\dots \lambda_{n-p}} \;
\E^{(n)}{}^{\mu_1 \dots \mu_{n},}{}_{\rho_1 \dots \rho_{p}}
{}^{\lambda_1\dots \lambda_{n-p}} = \E^{(n)}{\!}_{
\sigma_1 \dots \sigma_{p}\lambda_1\dots \lambda_{n-p},\,
\rho_1 \dots \rho_{p}} {}^{\lambda_1\dots \lambda_{n-p}} \nn \\
&\qquad {} = A^{(n)}_p  \,
\E^{(p)}{\!}_{\sigma_1 \dots \sigma_{p},\,\rho_1 \dots \rho_{p}} \, ,
\quad p = 0, \dots, n \, , \qquad
A^{(n)}_p =\frac{p!}{n!} \, \frac{\Gamma(d-p+1)}{\Gamma(d-n+1)} \, ,
\end{align}
and, if $n\ge1$,
\begin{align}
& \E^{(n)}{\!}_{\mu \, \mu_1 \dots \mu_{n-1},\,\sigma_1 \dots \sigma_{p}
\lambda_1\dots \lambda_{n-p}} \;
\E^{(n)}{\!}_\nu{\,}^{\mu_1 \dots \mu_{n-1},}{}_{\rho_1 \dots \rho_{p}}
{}^{\lambda_1\dots \lambda_{n-p}} \nn \\
&\quad {}= B^{(n)}_p \, \delta_{\mu}{\!}^{\lambda} \,\delta_{\nu}{\!}^{\lambda'}
\, \E^{(p)}{\!}_{\sigma_1 \dots \sigma_{p},\,\lambda \hskip 0.5pt
\lambda_1 \dots \lambda_{p-1}}
\; \E^{(p)}{\!}_{\rho_1 \dots \rho_{p},\,\lambda'}
{}^{\lambda_1 \dots \lambda_{p-1}}
+ C^{(n)}_p \, \eta_{\mu\nu} \, \E^{(p)} {\!}_{\sigma_1 \dots \sigma_{p},\,
\rho_1 \dots \rho_{p}} \, ,
\end{align}
where
\be
B^{(n)}_p =\frac{p\, p!}{n\, n!} \, \frac{\Gamma(d-p)}{\Gamma(d-n)} \, , \quad
C^{(n)}_p = \frac{(n-p)\, p!}{n\, n!} \,\frac{\Gamma(d-p)}{\Gamma(d-n+1)}
\, , \quad p=0,\dots, n\, .
\ee
Consistency requires $B^{(n)}_p + d \, C^{(n)}_p = A^{(n)}_p$,
$B^{(n)}_{p-1} = B^{(n)}_p \, B^{(p)}_{p-1}, \,
C^{(n)}_{p-1} = B^{(n)}_p \, C^{(p)}_{p-1} + B^{(n)}_p \, A^{(p)}_{p-1}$.
Evaluating \eqref{TAX} then gives
\be
C^{\rm gauge}_{T,n,2} = - \frac{2n(n+1)(n+3) \, (2n)!}{(n+2) \, n!^2} \, .
\label{Cgauge}
\ee
Note that $C^{\rm gauge}_{T,1,2} = C_{T,\vphi,4}$ for $d=4$.

Large $N$ methods, similar to those for the $O(N)$ and Gross--Neveu models,
have been extended to an Abelian gauge theory coupled to $N$
fermions~\cite{Kleb6}. For $d=4+2p$, $p=0,1,\dots$, this becomes equivalent
to a renormalisable theory with $N$ fermions and a higher-derivative gauge
theory with a Lagrangian $- \tfrac14 \, F^{\mu\nu}(-\pr^2)^p F_{\mu\nu}$.
Using the large $N$ results for $C_T$ and subtracting the free fermion
contribution~\cite{Kleb6}, in the notation used above, predicted that for
the free gauge theory $C^{\rm gauge}_{T,2,2p}= (-1)^p \,
2(p+2)(2p+4)!/((p+1)!(p+3)!)$. For $p=0$ this is the standard result and
the case $p=1$ agrees with \eqref{Cgauge} when $n=2$.

\section{\texorpdfstring{$(n-1)$}{(n-1)}-Form Theories away from Integer
Dimensions}

The usual gauge invariant action for $(n-1)$-form gauge fields is only
conformally invariant in $d=2n$ dimensions although it may be extended, as
in the previous section, in higher even dimensions with additional
derivatives. However abandoning gauge invariance the action may be extended
to be conformal for general $d$. The corresponding Weyl-invariant action on
a curved-space background was obtained by Erdmenger  \cite{Erdmenger},
following from the construction of a conformal second-order differential
operator on $k$-forms obtained by Branson \cite{Branson}, and the
corresponding flat-space action for vector fields or one-forms was given in
\cite{ElShowk}.

The curved-space action obtained in \cite{Erdmenger} may be expressed, with
a similar notation to \eqref{SA2}, as
\begin{align}
S_{n,0}[A]
= {}&- \frac{1}{2\,  n!} \,  \int \rmd^d x \sqrt{-\gamma} \; \Big (
F^{\mu_1 \dots \mu_{n}}  F_{\mu_1 \dots \mu_{n}}
+ \alpha \, \nabla_\lambda A^{\lambda \mu_1 \dots \mu_{n-2}} \,
 \nabla_\rho A^{\rho}{}_{\mu_1 \dots \mu_{n-2}}
 \nn  \\
\noalign{\vskip-6pt}
&\hskip 3 cm{}
+ \half n (d-2n)\big ( \gamma_{\lambda\rho} \, {\hat R} -
2(n-1) \, P_{\lambda\rho} \big )
A^{\lambda\mu_1 \dots \mu_{n-2}}   A^{\rho}{}_{\mu_1 \dots \mu_{n-2}} \Big )
\, , \label{SAz} \\
& \alpha = n(n-1) \, \frac{d-2n}{d-2n+4}  \, .  \nn
\end{align}
On flat space this is tantamount to a particular choice of a covariant
gauge fixing term \cite{ElShowk}.
The flat space energy-momentum tensor is then
\begin{align}
n! \,T_{n,0}^{\mu\nu} ={}& n\,
F^{\mu{\hskip 0.5pt} \mu_1 \dots \mu_{n-1}}  F^{\nu}{}_{\mu_1 \dots \mu_{n-1}}
- \half \, \eta^{\mu\nu} \,
F^{\mu_1 \dots \mu_n} F_{\mu_1 \dots \mu_n} \nn \\
&{} + (n-2)\alpha  \, \pr_\lambda
A^{\lambda \mu \hskip 0.5pt  \mu_1 \dots \mu_{n-3}}\,
\pr_\rho A^{\rho\hskip 0.5pt \nu}{}_{\mu_1 \dots \mu_{n-3}} - 2\alpha\,
 A^{(\mu \hskip 0.5pt  |\mu_1 \dots \mu_{n-2}}\, \pr^{\nu)}
\pr_\rho A^{\rho}{}_{\mu_1 \dots \mu_{n-2}} \nn \\
&{} - \half \alpha  \, \eta^{\mu\nu} \,  \pr_\lambda  A^{\lambda  \hskip 0.5pt  \mu_1 \dots \mu_{n-2}}\,
\pr_\rho A^{\rho}{}_{\mu_1 \dots \mu_{n-2}}
\nn   \\
&{}
-2(n-2)\alpha  \, \pr_\lambda \big (
A^{\lambda (\mu | \hskip 0.5pt  \mu_1 \dots \mu_{n-3}}\,
\pr_\rho A^{\rho\hskip 0.5pt |\nu)}{}_{\mu_1 \dots \mu_{n-3}} \big )
+ \alpha\, \eta^{\mu\nu} \, \pr_\lambda \big (
 A^{\lambda \hskip 0.5pt \mu_1 \dots \mu_{n-2}}\,
\pr_\rho A^{\rho}{}_{\mu_1 \dots \mu_{n-2}} \big )  \nn \\
 &{}
+ \half n(n-1) (d-2n)\,
\D^{\mu\nu\sigma\rho} \big ( A_\sigma{\hskip -0.5pt}{}^{\mu_1 \dots \mu_{n-2}}
A_{\rho{\hskip 0.5pt} \mu_1 \dots \mu_{n-2}}\big )
\nn \\
&{}- \frac{n(d-2n)}{4(d-1)} \, (\pr^\mu \pr^\nu - \eta^{\mu\nu} \pr^2 )
\big ( A^{\mu_1 \dots \mu_{n-1}}  A_{\mu_1 \dots \mu_{n-1}} \big ) \, .
\label{EM0}
\end{align}
This satisfies
\begin{align}
n! \, \pr_\mu T_{n,0}^{\mu\nu} = {}&
\big ( n \, \pr_\mu F^{\mu{\hskip 0.5pt} \mu_1 \dots \mu_{n-1}} +
\alpha \, \pr^{\mu_1} \pr_\rho A^{\rho \hskip 0.5pt \mu_2 \dots \mu_{n-1}}
\big ) F^\nu{\!}_{\mu_1 \dots \mu_{n-1}}
- \alpha \,  \pr^2 \pr_\rho A^{\rho \hskip 0.5pt\mu_1 \dots \mu_{n-2}}
A^{\nu}{}_{\mu_1 \dots \mu_{n-2}} \, ,
\nn \\
n! \, \eta_{\mu\nu} T_{n,0}^{\mu\nu} = {}& - \tfrac14(d-2n) \,
\big ( n \, \pr_\mu F^{\mu{\hskip 0.5pt} \mu_1 \dots \mu_{n-1}} +
\alpha \, \pr^{\mu_1} \pr_\rho A^{\rho \hskip 0.5pt \mu_2 \dots \mu_{n-1}}
\big ) A_{\mu_1 \dots \mu_{n-1}} \, ,
\label{Tn0}
\end{align}
and so the energy-momentum tensor is conserved and traceless on the
equations of motion. Of course for $d=2n$, $T_{n,0}^{\mu\nu}$ reduces
to the usual gauge invariant form.

The two-point function for the $(n-1)$-form field determined by the
action  \eqref{SAz}
on flat space was calculated in \cite{Erdmenger} by inverting the
Fourier transform of the kinetic differential operator and  then
returning to $x$-space giving
\begin{align}
\big \langle A_{\mu_1 \dots \mu_{n-1}} (x) & \,
A_{\nu_1 \dots \nu_{n-1}} (0) \big \rangle \nn \\
= {}&
\frac{ (n-1)!}{(d-2n)\, S_{d}} \, \frac{1}{(x^2)^{\frac12(d-2)}} \,
\E^{(n-1)}{\!}_{\mu_1 \dots \mu_{n-1},}{}^{\lambda_1 \dots \lambda_{n-1}}
I_{\lambda_1\, \nu_1}(x) \dots
I_{\lambda_n \, \nu_{n-1}}(x)  \, ,
\label{Atwo}
\end{align}
which has the form required by conformal invariance for
$ A_{\mu_1 \dots \mu_{n-1}} $ a conformal primary. From \eqref{Atwo}
\begin{align}
\big \langle  F_{\mu_1 \dots \mu_{n}} (x)  \,
A_{\nu_1 \dots \nu_{n-1}} (0) \big \rangle
= {}&
\frac{ n!}{(d-2n)\, S_{d}} \, \pr_\lambda
\bigg ( \frac{1}{(x^2)^{\frac12(d-2)}} \, \delta_\kappa{\!}^\eta
- 2(n-1) \, \frac{1}{(x^2)^{\frac12 d}} \, x_\kappa x^\eta \bigg ) \nn \\
\noalign{\vskip -4pt}
&{}\hskip 2cm {}\times
\E^{(n)}{\!}_{\mu_1 \dots \mu_{n-1},}{}^{\lambda\hskip 0.5pt
\kappa \hskip 0.5pt\lambda_1 \dots
\lambda_{n-2}} \, \E^{(n-1)}{\!}_{\eta \lambda_1 \dots \lambda_{n-2},\,
\nu_1 \dots \nu_{n-1}} \nn \\
={}& - \frac{n!}{S_{d}} \,
\frac{1}{(x^2)^{\frac12 d }} \, x^\lambda \,
\E^{(n)}{\!}_{\mu_1 \dots \mu_{n},\,\lambda \hskip 0.8pt \nu_1 \dots \nu_{n-1}} \, ,
\label{FAdtwo}
\end{align}
and
\begin{align}
\big \langle  F_{\mu_1 \dots \mu_{n}}& (x)  \,
F_{\nu_1 \dots \nu_{n}} (0) \big \rangle  =
\frac{ n\,n!}{S_{d}} \, \pr_\rho
\bigg ( \frac{1}{(x^2)^{\frac12 d}} \, x^\lambda \bigg )
\E^{(n)}{\!}_{\mu_1 \dots \mu_{n-1},\,\lambda\hskip 0.5pt\lambda_1 \dots
\lambda_{n-1}} \, \E^{(n)}{}^{\rho \hskip 0.5pt \lambda_1 \dots \lambda_{n-1},}
\,{}_{\nu_1 \dots \nu_{n}} \nn \\
={}& \frac{ n\,n!}{S_{d}} \,
\frac{1}{(x^2)^{\frac12 d }} \bigg ( \delta_\lambda{\!}^\rho
- d \, \frac{1}{x^2 } \, x_\lambda x^\rho \bigg ) \,
\E^{(n)}{\!}_{\mu_1 \dots \mu_{n},}{}^{\lambda\hskip 0.5pt
\lambda_1 \dots \lambda_{n-1}} \,
\E^{(n)}{\!}_{\rho \lambda_1 \dots \lambda_{n-1},\, \nu_1 \dots \nu_{n}} \, .
\label{Fdtwo}
\end{align}
This has the conformally invariant form in terms of the inversion tensor
only when $d=2n$ and is identical in this case to the expression obtained
for gauge choices for $A_{\mu_1 \dots \mu_{n-1}}$ other than that
implicit in \eqref{Atwo} \cite{CFTNotes}.

As in other cases, evaluating the energy-momentum tensor two-point function
can be simplified to
\begin{align}
& \big \langle T_{n,0}^{\hskip 1pt \mu\nu}(x) \,
T_{n,0}^{\hskip 1pt \sigma\rho}(0)  \big \rangle =
\big \langle T_{n,0}^{\hskip 1pt \mu\nu}(x) \, Y_n^{\sigma\rho}(0)
\big \rangle \, ,  \nn \\
& n! \,Y_n^{\sigma\rho} = n\,
F^{\sigma{\hskip 0.5pt} \mu_1 \dots \mu_{n-1}}  F^{\rho}{}_{\mu_1 \dots \mu_{n-1}}
 + (n-2)\alpha  \, \pr_\lambda
A^{\lambda \sigma \hskip 0.5pt  \mu_1 \dots \mu_{n-3}}\,
\pr_\kappa A^{\kappa\hskip 0.5pt \rho}{}_{\mu_1 \dots \mu_{n-3}} \nn \\
\noalign{\vskip -3pt}
&\hskip 1.6cm {} - 2\alpha\,
 A^{(\sigma \hskip 0.5pt  |\mu_1 \dots \mu_{n-2}}\, \pr^{\rho)}
\pr_\kappa A^{\kappa}{}_{\mu_1 \dots \mu_{n-2}} \, .
\label{TYX}
\end{align}
This then determines \be C_{T,n} = \frac{d}{d-1} \,
\frac{(d-n+2)_{n-1}}{(n-1)!} \, , \qquad n=1,2,\dots \, .  \label{CTn} \ee
As expected $C_{T,1}=C_{T,S}$. The corresponding result for $(n-1)$-form
gauge fields in $d=2n$ dimensions, whose energy-momentum tensor is obtained
just from the $FF$ terms in \eqref{Tn0}, is $C^{\rm gauge}_{T,n,0} = 2\,
n^2 (2n-2)!/(n-1)!^{\hskip 0.5pt 2}$, \cite{Buchel, CFTNotes}. This is not
equal to $C_{T,n}$ in \eqref{CTn} when $d=2n$, although \eqref{EM0}
apparently reduces to the required form for this $d$.\footnote{If the
energy-momentum  tensor \eqref{EM0} is restricted to the gauge-invariant
$FF$ terms and we use \eqref{Fdtwo} then the resulting two-point function
is not of the required conformal form \eqref{Tdef}. If $C_T$ is identified
through the coefficient of the $xxxx$ terms for $n=2$ then $C_{T,2}= \half
d^2 (d-2)$ as obtained in \cite{Kleb6}. This prescription in general gives
$C_{T,n}=\half\, d^2 (d-2)\cdots (d-n)/(n-1)!$.} The difference arises
since the $\langle A A \rangle$ two-point function in \eqref{Atwo} is also
singular when $d=2n$.  The representation of the conformal group generated
from a conformal primary $A_{\mu_1 \dots \mu_{n-1}}$ is reducible when
$d=2n$ and an irreducible representation for the associated gauge theory is
obtained by  quotienting by the invariant subspace corresponding to gauge
transformations. Since $C_T$ is related to the number of  degrees of
freedom it is expected to differ between the gauge theory and that
corresponding to the $(n-1)$-form  $A_{\mu_1 \dots \mu_{n-1}}$.

To demonstrate this further we may consider the Fourier transform of the
two-point function in \eqref{Atwo} letting $\half(d-2)\to \Delta, \, k=n-1$
and also setting the overall coefficient to 1,
\begin{align}
& \pi^{\frac12 d} \, \frac{\Gamma(\Delta-\half d+1)}{\Gamma(\Delta+1)} \, F(p^2) \,
A_{\mu_1 \dots \mu_{k},\, \nu_1 \dots \nu_{k}} (p) \, , \qquad
F(p^2 ) = - \frac{\pi} { \sin \pi(\Delta - \half d)} \, \big (\tfrac14 \, p^2 \big )^{\Delta - \frac12 d} \, ,
\nn \\
& A_{\mu_1 \dots \mu_{k},\, \nu_1 \dots \nu_{k}} (p) =
\bigg ( (\Delta - k) \, \delta_\lambda{\!}^\rho
- 2 k(\Delta - \half d)  \, \frac{1}{p^2 } \, p_\lambda\, p^{\hskip 0.8pt \rho} \bigg ) \,
\E^{(k)}{\!}_{\mu_1 \dots \mu_{k},}{}^{\lambda\hskip 0.5pt
\lambda_1 \dots \lambda_{k-1}} \,
\E^{(k)}{\!}_{\rho \lambda_1 \dots \lambda_{k-1},\, \nu_1 \dots \nu_{k}} \, .
\end{align}
Noting that $\big (F(p^2-i\epsilon) - F(p^2+i \epsilon)\big )/2\pi i =
\theta(-p^2) \,  \big ({-\tfrac14}\, p^2 \big )^{\Delta - \frac12 d} $
unitarity requires that the matrix $A_{\mu_1 \dots \mu_{k},\, \nu_1 \dots
\nu_{k}} (p) $ should be positive definite for $p^2<0$. The eigenvalues
are $\Delta-k$, $d-\Delta-k$, but for the the second case the eigenvectors
$p_{[\mu_1}\, \epsilon_{\mu_2 \dots \mu_{k}]}$ have negative norm for
$p^2<0$ so that we must have, for a unitary CFT of $k$-forms,
\be \Delta > k
\, , \qquad \Delta> d- k \, .
\label{posA}
\ee
When $\Delta=k$ or $\Delta=d-k$ there are zero modes related to the
reducibility of the representation. For the case of interest above
$\Delta=\half(d-2)$ and $\Delta=k$ corresponds to $d=2n$. For $d$ an integer
it should be noted that the conditions \eqref{posA}  are invariant under
duality $A_{\mu_1 \dots \mu_k} \to ({}^*A)_{\mu_1 \dots \mu_{d-k}}$.

\section{Conformal Primary Operators}
The energy-momentum tensor is a conformal primary operator. The detailed
expressions in \eqref{T4phi} and \eqref{T6phi} are necessary to ensure this
and we show here how they can be recovered by requiring
$T_{\vphi,n,\mu\nu}$ to be a conformal primary, and that this determines
the parameter $\lambda$ in accord with \eqref{defl}, although this term is
conserved and traceless by itself.

For a local tensor operator $X_{\alpha_1 \dots \alpha_n}$ formed from
multinomials in $\vphi$ and derivatives at $x=0$ we define
\eqn{\big [ K_b , \pr_\mu \big ] X_{\alpha_1 \dots \alpha_n} = b_\mu
\, D \,X_{\alpha_1 \dots \alpha_n}  + {\ts{\sum_i}} \; \big (
b_{\alpha_i} \, X_{\alpha_1 \dots\mu \dots \alpha_n}
- \eta_{\mu \alpha_i} \, b^\lambda
X_{\alpha_1 \dots \lambda \dots \alpha_n} \big ) \, ,
\label{actK}}[]
with
\eqn{D \, \pr_\mu = \pr_\mu ( D+1) \, , \qquad
D \, X_{\alpha_1 \dots \alpha_n} = \Delta_X \,
X_{\alpha_1 \dots \alpha_n} \, ,}[]
where $\Delta_X$ is determined just by counting the number of derivatives
and fields $\vphi$ in $ X_{\alpha_1 \dots \alpha_n}$. For any conformal
primary $X_A$, $A=\{\alpha_1 \dots \alpha_n \} $,  then $K_b \, X_{A} = 0$
(this is of course the usual condition $K_\mu X_A(0)=0$).
Otherwise $X_{A}$ is not a conformal primary and generates a reducible
representation of the conformal group.  Acting with $K_b$ removes
derivatives so that for some finite $N$, $K_b{\!}^{N+1}  X_{A} = 0$ and
hence we can write $K_b{\!}^{N}  X_{A} = \sum_{r,I} f_{A,r I} (b)  \, O_I$,
where $\{O_I\}$ is a basis of conformal primaries with $\Delta_{O_I} =
\Delta_X -N$ and $ f_{A,r I} (b) = {\rm O}(b^N)$. For $Y_A =  \sum_{r,I}
\D_{A,r I} (\pr) \, O_I$ then $K_b{\!}^{N}  (X_{A} - Y_A)=0$ gives
$\sum_{s,J} M_{rI, sJ} \,  \D_{A,s J} (b) = f_{A,r I} (b)$.  Extending $\{
f_{A,r I} (b) \} $ to include all possible rotationally covariant forms,
with $f_{A,r I}=0$ for some $r$ if necessary, $M$ is a square matrix and we
may solve for $ \D_{A,r I} (b)$ unless $\{ O_I\}$ have conformal primary
descendants with $N$ derivatives so that $\det M=0$. For generic $\Delta_X$
this does not arise. Iterating this construction then gives the conformal
primary $X_A - \sum_Y Y_A$ which is the lowest weight state for an
irreducible representation. If $\det M=0$, for particular $\Delta_X$,
the representation space obtained
from $X_A$ is reducible but not decomposable.

The result \eqref{actK} can be extended successively to multiple
derivatives. For a scalar conformal primary $\vphi$ with scale dimension
$\delta$, so that $D\, \vphi = \delta \, \vphi$, and an arbitrary vector
$a$, we have
\eqn{K_b \, (a\cdot \pr)^n \vphi   = n (\delta  +n-1)  \, a \cdot b \,
(a\cdot \pr)^{n-1} \vphi
- \half n(n-1) \, a^2 \, b\cdot \pr \,  (a\cdot \pr)^{n-2} \vphi \, ,
\label{Knn}}[]
from which, by acting with $\pr_a \cdot \pr_a$,
\eqna{K_b \, (a\cdot \pr)^{n} \pr^2 \vphi  = {}&
n (\delta  +n + 1)  \, a \cdot b \,  (a\cdot \pr)^{n-1} \pr^2 \vphi
+ (2\delta  +2 - d) \, b\cdot \pr \, (a\cdot \pr)^n  \vphi  \\
&{}
- \half n(n-1) \, a^2 \, b\cdot \pr \,  (a\cdot \pr)^{n-2} \pr^2 \vphi \, .
}[]
From \eqref{Knn} then
\eqn{\Phi_n(a)= a^{\mu_1} \dots a^{\mu_n} \, \Phi_{\mu_1 \dots \mu_n} =
\sum_{r=0}^n \binom{n}{r} \, \frac{(-1)^r}{(\delta)_r \, (\delta)_{n-r}} \,
(a\cdot \pr)^r \vphi \; (a\cdot \pr)^{n-r} \vphi
\label{Phin}}[]
satisfies $ K_b \,\Phi_n(a) = {\rm O}(a^2)$ and so taking $a^2=0 $, which
projects out the traces, this demonstrates that $\Phi_n(a)$ defines a
symmetric traceless conformal primary with $\Delta_{\Phi_n} = 2\delta+n$
and twist $2\delta$ \cite{Mikhailov, Braun}. For higher twist results are
more complicated. In the following we will work out $\text{O}(a^2)$ terms
in a few examples and obtain some results for higher twist. Note that
\eqref{Phin} of course gives $\Phi_n(a)=0$ for $n$ odd.

In the remainder of this section we apply the procedure outlined above for
constructing conformal primaries.  Initially we construct a conformal
primary starting from $\pr_\mu \vphi \, \pr_\nu \vphi$. Using \eqref{actK}
we get
\eqn{K_b{\!}^2 (\pr_\mu \vphi \, \pr_\nu \vphi) = 2 \delta^2\, b_\mu b_\nu \,
\vphi^2 \, ,}[]
and also
\eqn{K_b{\!}^2 (\pr_\mu \pr_\nu \vphi^2)  = 2 \delta \big ( 2(2\delta+1) \,
b_\mu b_\nu  - \eta_{\mu\nu} \, b^2 \big ) \vphi^2 \, .}[]
Hence, a symmetric tensor conformal primary with dimension $\Delta_{O_{2}}
= 2 \delta +2$ and twist $2\delta$ is given by
\eqna{O_{2,\mu\nu} = {}& \pr_\mu \vphi \, \pr_\nu \vphi -
\frac{\delta}{2(2\delta+1)} \Big ( \pr_\mu\pr_\nu
+ \frac{1}{4\delta + 2 -d} \, \eta_{\mu\nu} \, \pr^2 \Big ) \vphi^2  \\
= {}& - \pr_\mu \pr_\nu \vphi \, \vphi +
\frac{1}{2(2\delta+1)} \Big ( (\delta+1)\, \pr_\mu\pr_\nu
- \frac{\delta}{4\delta + 2 -d} \, \eta_{\mu\nu} \, \pr^2 \Big ) \vphi^2 \, ,
\label{Otwo}}[]
so that $K_b\,O_{2,\mu\nu}=0$. A scalar conformal primary is then
\eqn{\eta^{\mu\nu}O_{2,\mu\nu}=-\pr^2\vphi\,\vphi+\frac{2\delta+2-d}
{2(4\delta+2-d)}\,\pr^2\vphi^2\,.}[]
For $\delta = \half(d-2)$ we have
\eqn{T_{\vphi,2,\mu\nu} = O_{2,\mu\nu}  - \half \,
\eta_{\mu\nu} \, O_{2,\sigma\sigma}\,.}[]

With more derivatives the construction becomes more lengthy as the
descendants of more conformal primaries have to be subtracted.  To
construct conformal primary symmetric tensors with four derivatives we
start from
\eqn{K_b{\!}^4 \big ( (a\cdot \pr)^4  \vphi \, \vphi\big ) =
6 \delta (\delta+1) \big ( 4 (\delta+2)(\delta+3) \, (a\cdot b) ^4
- 12 (\delta+2)
\, (a\cdot b)^2 a^2 b^2 + 3 (a^2 b^2)^2 \big ) \vphi^2 \, .
\label{Kbfour}}[]
This can be cancelled by terms involving four derivatives acting on
$\vphi^2$ so that
\eqna{K_b{\!}^2 \big ( (a\cdot \pr)^4  \vphi \, \vphi - \D_{4,\vphi^2} \, \vphi^2 \big )&=
- 6 (\delta+2) \big ( 2 (\delta+3) (a\cdot b)^2 - a^2 b^2 \big )
O_{2,aa} \\
&\quad + 24(\delta+2) \, a\cdot b \, a^2 \,  O_{2,ab}
+ 6 (a^2)^2 \, O_{2,bb} \, ,
\label{Kb2}}[]
where
\eqna{\D_{4,\vphi^2} = {}&  \frac{1}{4(2\delta+1)(2\delta+3)} \Big ( (
\delta+2)(\delta+3) \, (a\cdot \pr)^4
- \frac{6\delta(\delta+2)}{4\delta+2-d} \, a^2 (a\cdot \pr)^2 \pr^2 \\
\noalign{\vskip-4pt}
& \hskip 5cm
{} + \frac{3\delta(\delta+1)}{(
4\delta+2-d)(4\delta+4-d)} \, (a^2)^2 (\pr^2)^2 \Big )  \, ,
\label{Dfourp}}[]
and $O_{2,aa}= a^\mu a^\nu O_{2,\mu\nu}$ with $O_{2,\mu\nu}$ the conformal
primary given by \eqref{Otwo}. By adding extra contributions  with two
derivatives acting on $O_{2,\mu\nu}$ the remaining terms in \eqref{Kb2} may
be cancelled so as to obtain a four index conformal primary
\eqn{a^\mu a^\nu  a^\sigma a^\rho  \, O_{4,\mu \nu \sigma\rho}
=   (a\cdot \pr)^4  \vphi \, \vphi   - \D_{4,\vphi^2}\,  \vphi^2 - \D_4{}^{\mu\nu}
O_{2,\mu\nu} \, ,}[]
where
\eqna{\D_4{}^{\mu\nu} = & {}- \frac{3}{2\delta+5} \Big ( (\delta+3)
(a\cdot \pr)^2
- \frac{\delta+2}{4\delta+6-d} \, a^2 \,
\pr^2 \Big ) \, a^\mu a^\nu \\
&{} - \frac{3}{(2\delta+5)(4\delta+6-d)} \Big ( 2 \, a^2 a\cdot \pr \,
a^{(\mu} \pr^{\nu)}  + \frac{1}{2\delta+3-d} \,
(a^2)^2 \, \pr^\mu \pr^\nu \Big )
 \\
&{} +
\frac{3}{(2\delta+3)(2\delta+5)(4\delta+6-d)}
\Big (  a^2 (a\cdot \pr)^2  +
\frac{1}{2\delta+3-d} \,  (a^2)^2 \pr^2 \Big ) \, \eta^{\mu\nu} \, .
\label{Dfour2}}[]
It is straightforward to check that $K_b\, O_{4,\mu\nu\rho\sigma}=0$, as
guaranteed by the fact that there are no primaries with three derivatives
and two $\vphi$'s. By acting with $\pr_a \cdot \pr_a$ we obtain a spin-two
primary with twist $2\delta+2$,
\eqna{\eta^{\sigma\rho} a^\mu a^\nu &  \, O_{4,\mu \nu \sigma\rho}  \\
= {}&  (a\cdot \pr)^2 \pr^2 \vphi \, \vphi \\
&{} - \frac{2\delta+2-d}{4(2\delta+1)(4\delta+2-d)}
\Big ( (\delta+2)\, (a\cdot \pr)^2 \pr^2
-\frac{\delta }{4\delta+4-d}
 \, a^2 \, (\pr^2)^2 \Big ) \vphi^2   \\
& {}+ \frac{1}{2(4\delta+6-d)} \Big (
(2\delta+2-d) \, \pr^2 \, a^\mu a^\nu \\
\noalign{\vskip -6pt}
&\hskip 3cm {} + 2(4\delta+8-d)  \, a\cdot \pr \,
a^\mu \pr^\nu  + \frac{2}{2\delta+3-d} \,
a^2 \, \pr^\mu \pr^\nu \Big ) O_{2,\mu\nu}
 \\
&{} +
\frac{1}{2(2\delta+3)(4\delta+6-d)}
\Big (  \big ( (2(\delta+1)(2\delta+5) - d(\delta+2) \big ) \,
(a\cdot \pr)^2
 \\
&\hskip 5cm {} -
\frac{ 2\delta^2 + 5\delta +5 - d(\delta+1)}{2\delta+3-d} \,
a^2 \pr^2 \Big ) \,  \eta^{\mu\nu} O_{2,\mu\nu} \, ,
\label{Ofour2}}[]
as well as a scalar primary,
\eqna{\eta^{\mu\nu}\eta^{\sigma\rho} O_{4,\mu \nu \sigma\rho} = {}&
(\pr^2)^2 \vphi \, \vphi
- \frac{(\delta+1)(2\delta+2-d)(2\delta+4-d)}
{2(2\delta+1)(4\delta+2-d)(4\delta+4-d)}
\, (\pr^2)^2  \vphi^2   \\
& {}+ \frac{2\delta+4-d}{2\delta+3-d} \, \Big (  \pr^\mu \pr^\nu
+ \frac{2\delta+2-d}{4\delta+6-d} \;
 \pr^2  \,  \eta^{\mu\nu} \Big ) O_{2,\mu\nu} \, .
\label{Ofour3}}[]
For $\delta = \half(d-4)$,
\eqn{T_{\vphi,4,\mu\nu} = 2\,O_{4,\mu\nu\sigma\sigma} -
\half \, \eta_{\mu\nu}\, O_{4,\sigma\sigma\rho\rho}\,.}[]
In this case $ 4 \delta+6-d= d-2$ and the construction of $T_{4,\mu\nu}$
fails when $d=2$ since then $O_{2,\mu\nu}$ has a spin two conformal primary
descendant with two derivatives.

For six derivatives \eqref{Kbfour} is extended to
\eqna{K_b{\!}^6 \big ( (a\cdot \pr)^6  \vphi \, \vphi\big ) =
90 \delta (\delta+1)(\delta+2) & \big ( 8 (\delta+3)(\delta+4)(\delta+5)
\, (a\cdot b) ^6  \\
&{} - 60 (\delta+3)(\delta+4)
\, (a\cdot b)^4 a^2 b^2  \\
&{} + 90 (\delta+3)  (a\cdot b)^2 (a^2 b^2)^2
- 15 (a^2 b^2)^3 \big ) \vphi^2 \, .
\label{Kbsix}}[]
Then,
\eqna{K_b{\!}^4 \big (& (a\cdot \pr)^6  \vphi \, \vphi
- \D_{6,\vphi^2} \, \vphi^2 \big )  \\
= {}& -90 (\delta+2) (\delta+3)\Big ( 4 (a\cdot b)^2
(\delta+4)  \big ( (\delta+5)
(a\cdot b)^2 - 3 \,  a^2 b^2 \big ) O_{2,aa}    \\
&\hskip 3.5cm {} + 3 (a^2 b^2)^2  O_{2,aa}
- 8 \, a\cdot b \, a^2
\big ( 2 (a\cdot b)^2 - 3\,  a^2 b^2 \big ) O_{2,ab}  \\
&\hskip 3.5 cm {} + 6 (a^2)^2 \big ( 2 (\delta+3)(a\cdot b)^2 -
a^2 b^2 \big ) O_{2,bb} \Big ) \, ,
\label{Kb4}}[]
with
\eqna{\D_{6,\varphi^2}=\frac{1}{8(2\delta+5)}
\bigg(&(\delta+3)(\delta+4)\bigg(\frac{\delta+5}{(2\delta+1)
(2\delta+3)}\,(a\cdot\pr)^2\\
&\phantom{(\delta+3)(\delta+4)\Bigg(}
-\frac{15\delta}{\big(4\delta(\delta+2)+3\big)
(4\delta+2-d)}\,a^2\pr^2\bigg)(a\cdot\pr)^4\\
&+\frac{15\delta(\delta+1)}{(4\delta+2-d)(4\delta+4-d)}
\bigg(\frac{3(\delta+3)}{4\delta(\delta+2)+3}\,
(a\cdot\pr)^2\\
&\phantom{-\frac{15\delta(\delta+1)}{(4\delta+2-d)(4\delta+4-d)}
\bigg(}-\frac{\delta+2}{(2\delta+1)(2\delta+3)(4\delta+6-d)}\,a^2\pr^2\bigg)(a^2\pr^2)^2\bigg).
\label{Dsixp}}[]
Further, we have
\eqna{K_b{\!}^2 \big (&(a\cdot \pr)^6  \vphi \, \vphi
- \D_{6}{}^{\mu\nu} \,  O_{2,\mu\nu}
- \D_{6,\vphi^2}  \, \vphi^2 \big )  \\
= {}& 15 (\delta+4)  \big ( 2 (\delta+5)
(a\cdot b)^2 -  a^2 b^2 \big ) O_{4,aaaa}
-120(\delta+4)\,  {a\cdot b} \, a^2 O_{4,aaab}
+90 (a^2)^2\,  \O_{4,aabb} \, ,
\label{Kb6}}[]
with
\eqna{\D_6{}^{\mu\nu}&=-\frac{15}{4(2\delta+5)(2\delta+7)}
\Big(
(\delta+4)(\delta+5)(a\cdot\pr)^4
  -\frac{6(\delta+2)(\delta+4)}{4\delta+6-d}\,a^2(a\cdot\pr)^2\pr^2\\
  &\hspace{6.5cm}+\frac{3(\delta+2)(\delta+3)}{(4\delta+6-d)
  (4\delta+8-d)}(a^2)^2(\pr^2)^2
  \Big) a^\mu a^\nu\\
  &\quad-\frac{45}{(2\delta+5)(2\delta+7)(4\delta+6-d)}\Big(
  (\delta+4)(a\cdot\pr)^2
  -\frac{\delta+2}{4\delta+8-d}\,a^2\pr^2\Big)a^2a\cdot\pr\,
  a^{(\mu}\pr^{\nu)}\\
  &\quad-\frac{45}{2(2\delta+5)(2\delta+7)(2\delta+3-d)(4\delta+6-d)
  (4\delta+8-d)}\\
  &\hspace{4cm}\times\Big(
  \big(4\delta^2+26\delta+36-
  d(\delta+5)\big)(a\cdot\pr)^2-(\delta+2)\,a^2\pr^2\Big)
  (a^2)^2\pr^\mu\pr^\nu\\
  &\quad+\frac{45}{2(2\delta+3)(2\delta+5)(2\delta+7)(4\delta+6-d)}\Big(
  (\delta+4)(a\cdot\pr)^4\\
  &\quad\phantom{+\frac{45}{2(2\delta+3)(2\delta+5)(2\delta+7)(4\delta+6-
  d)}\Big(}+\frac{2\delta^2+19\delta+30-3d}{(2\delta+3-d)(4\delta+8-
  d)}\,a^2(a\cdot\pr)^2\pr^2\\
  &\quad\phantom{+\frac{45}{2(2\delta+3)(2\delta+5)(2\delta+7)(4\delta+6
  -d)}\Big(}-\frac{\delta+2}{(2\delta+3-d)(4\delta+8-d)}\,(a^2)^2(\pr^2)^2
  \Big)a^2\eta^{\mu\nu}\,.
\label{Dsix2}}[]
A two-derivative operator on $O_{4,\mu\nu\sigma\rho}$ can now
be constructed to obtain a conformal primary with six indices, namely
\eqn{a^\mu a^\nu a^\sigma a^\rho a^\tau a^\omega
\,O_{6,\mu\nu\sigma\rho\tau\omega}=-(a\cdot \pr)^6\vphi\,\vphi
+\D_{6,\vphi^2}\,\vphi^2 + \D_6{}^{\mu\nu} O_{2,\mu\nu}
+\D_6{}^{\mu\nu\sigma\rho}\,O_{4,\mu\nu\sigma\rho} \, ,
}[]
where
\eqna{\D_6{}^{\mu\nu\sigma\rho}&=\frac{15}{2(2\delta+9)}\Big((\delta+5)(a\cdot\pr)^2-\frac{\delta+4}{4\delta+10-d}a^2\pr^2\Big)a^\mu
a^\nu a^\sigma a^\rho\\
&\quad+\frac{30}{(2\delta+9)(4\delta+10-d)}\,a\cdot\pr\,a^{(\mu}a^\nu
a^\sigma \pr^{\rho)}\\
&\quad+\frac{45}{(2\delta+9)(2\delta+3-d)(4\delta+10-d)}\,(a^2)^2 a^{(\mu}
a^\nu\pr^\sigma\pr^{\rho)}\\
&\quad-\frac{45}{(2\delta+7)(2\delta+9)(4\delta+10-d)}\Big(
(a\cdot\pr)^2+\frac{1}{2\delta+3-d}\,a^2\pr^2\Big)a^{(\mu} a^{\nu}
\eta^{\sigma\rho)}\\
&\quad-\frac{180}{(2\delta+7)(2\delta+9)(2\delta+3-d)(4\delta+10-d)}\,
(a^2)^2 a\cdot\pr\,a^{(\mu}\pr^{\nu}\eta^{\sigma\rho)}\\
&\quad-\frac{90}{(2\delta+7)(2\delta+9)(2\delta+3-d)(2\delta+5-d)
(4\delta+10-d)}\,(a^2)^3\pr^{(\mu}\pr^{\nu}\eta^{\sigma\rho)}\\
&\quad+\frac{90}{(2\delta+5)(2\delta+7)(2\delta+9)(2\delta+3-d)
(4\delta+10-d)}\\
&\hspace{5cm}\times\Big((a\cdot\pr)^2
+\frac{1}{2\delta+5-d}a^2\pr^2\Big)(a^2)^2\eta^{\mu(\nu}\eta^{\sigma)\rho}\,.
\label{Dsix4}}[]
We have $K_b\,O_{6,\mu\nu\sigma\rho\tau\omega}=0$ since there are no
conformal primaries with five derivatives and two $\vphi$'s.

A four-index as well as a two-index and a scalar conformal primary can be
obtained from $O_6$ by acting with $\pr_a\cdot\pr_a$.  To avoid even more
lengthy expressions we only list here the scalar primary,
\eqna{\eta^{\mu\nu}\eta^{\sigma\rho}\eta^{\tau\omega}O_{6,\mu\nu\sigma
\rho\tau\omega}&=-(\pr^2)^3\vphi\,\vphi\\
&\quad+\frac{(\delta+2)(2\delta+2-d)(2\delta+4-d)(2\delta+6-d)}
{4(2\delta+1)(4\delta+2-d)(4\delta+4-d)(4\delta+6-d)}\,(\pr^2)^3\vphi^2\\
&\quad-\frac{3(2\delta+4-d)(2\delta+6-d)}{2(2\delta+3-d)
(4\delta+6-d)}\Big(\pr^\mu\pr^\nu\\
&\quad\phantom{+\frac{3(2\delta+4-d)(2\delta+6-d)}{2(2\delta+3-d)
(4\delta+6-d)}\Big(}+\frac{2\delta^2+5\delta+1-(\delta+2)d}
{(2\delta+3)(4\delta+8-d)}\,\pr^2\,\eta^{\mu\nu}\Big)\,\pr^2 O_{2,\mu\nu}\\
&\quad+\frac{3(2\delta+6-d)}{2(2\delta+5-d)}\Big(2\,\pr^\mu\pr^\nu
+\frac{2\delta+3-d}{4\delta+10-d}\,\pr^2\,\eta^{\mu\nu}\Big)
O_{4,\mu\nu\sigma\sigma}\,.
}[]
For $\delta=\frac12(d-6)$ we can express
\eqn{T_{\vphi,6,\mu\nu}=3\,O_{6,\mu\nu\sigma\sigma\rho\rho}
-\tfrac12\,\eta_{\mu\nu}\,O_{6,\tau\tau\sigma\sigma\rho\rho}\,,\qquad
\text{if}\quad \lambda=-\frac{8}{d-4}\,.}[]
Thus, we see that the requirement that $T_{\vphi,6,\mu\nu}$ be a conformal
primary determines $\lambda$, independently and consistently with the
result \eqref{defl} obtained from the curved-space action contribution of
the Bach tensor.

The poles in \eqref{Dfourp}, \eqref{Dfour2}, \eqref{Dsixp}, \eqref{Dsix2}, \eqref{Dsix4}
at $4\delta = 4 - 2k$,  for $k=1,2,\dots $, arise for these $\delta$ since there are
corresponding  differential operators
generating conformal primary descendants given by, for $O_{\mu_1 \dots \mu_\ell}$
a symmetric traceless tensor,
\begin{align}
\D_{n,\ell}^{\mu_1 \dots \mu_\ell} O_{\mu_1 \dots \mu_\ell}
= {}& \sum_{r=0,-n}^\ell  (-1)^r
\frac{(\tfrac12 d + \ell+n-1)_r}{2^r\, r!\, (\ell-r)!\, (n+r)!} \,
  (a\cdot \pr)^{\ell-r} (\pr^2)^{n+r} a^{\mu_1} \dots a^{\mu_r}
\pr^{\mu_{r+1}} \pr^{\mu_\ell} \, O_{\mu_1 \dots \mu_\ell}\, ,  \nn  \\
& a^2=0, \quad n+\ell \ge 1, \quad \Delta_O= \half d - n-\ell  \, ,
\end{align}
is a conformal
primary symmetric traceless tensor of rank $\ell$ and $\Delta = \half d + n$. Thus
$(\pr^2)^n O$ is a conformal primary scalar for $\Delta_O = \half d -n$. The
poles \eqref{Dfour2}, \eqref{Dsix2}, \eqref{Dsix4} at $2\delta = d - n$ for $n=3,5$ correspond
to conformal primary descendants
\be
\pr^{\mu_{k}}\dots  \pr^{\mu_\ell} \, O_{\mu_1 \dots \mu_\ell} \, , \quad
k = 1, \dots , \ell  \, , \quad
\Delta_O = d +k -2 \, ,
\ee
which are symmetric traceless tensors of rank $\ell-k$. There are also conformal
primary traceless tensor descendants of the form
\be
(a\cdot \pr)^k  a^{\mu_{1}}\dots  a^{\mu_\ell} \, O_{\mu_1 \dots \mu_\ell} \, , \quad
a^2=0 \, , \quad k = 1,2,\dots \, , \quad \Delta_O = 1 - \ell-k \, ,
\ee
of rank $k+\ell$ and correspond to the poles at $2\delta = -n$ for $n=1,3,\dots$.

\section{Conclusion}

In this paper we have computed $C_T$ for various free field theories
outside the usual range. Although such theories involving higher
derivatives  in general correspond to non-unitary quantum field theories,
they appear to be relevant in understanding some CFTs for large $N$ numbers
of component fields where the $1/N$ expansion remains valid for arbitrary
dimension $d$. Of course there are additional parameters, such as those
associated with the energy-momentum tensor three-point function, one of
which is related to $C_T$ by Ward identities. For the usual free CFTs these
were also calculated in \cite{Pet}.  The theories discussed here might also
be extended  to determine the energy-momentum tensor three-point function,
but the complexity of the expressions for $T_{\mu\nu}$ makes this a rather
formidable task, as are the corresponding large $N$ calculations.

In general in even dimensions $C_T$ in a CFT is related to  a particular
term quadratic in the Weyl tensor in the energy-momentum tensor trace on a
curved space background.  In four and six dimensions, where the trace
anomaly coefficients are $c$ and $c_3$,  the relations are \cite{6dimP}
\be
C_T = \tfrac43 \times 5! \, c \, , \qquad C_T = \tfrac35 \times 7! \, c_3 \, ,
\ee
with a normalisation chosen so  that for conventional free field theories
$c,c_3$ are given by $5!\, c= n_S + 3\, n_W + 12 \, n_A$, $7! \, c_3 = 2 \,
n_S + 40\, n_W + 180\, n_B$ with $n_S$ scalars, $n_W$ Weyl fermions and
$n_A,n_B$ the number of vector, $2$-form gauge fields in four, six
dimensions. There is of course complete agreement between the results for
$C_T$ and the curved space results based on using the heat kernel for
second-order conformal differential operators.  The results obtained here
allow some contributions to the heat kernel for higher-order operators on
curved backgrounds to be obtained. These have been discussed for Ricci flat
backgrounds in~\cite{Beccaria1,Beccaria2}.

Heat kernel techniques allow a perturbation expansion for arbitrary curved
backgrounds so as to determine the leading corrections to $c,c_3$. In six
dimensions for a cubic interaction $\tfrac16 \, \lambda_{ijk} \phi_i \phi_j
\phi_k$ then results in \cite{Grinstein:2015ina, 6dimP} give to lowest
order
\be
7! \, c_{3,1}
= - \tfrac{7}{36} \, {\hat \lambda}_{ijk}{\hat \lambda}_{ijk} \, ,
\qquad {\hat \lambda}_{ijk} = \lambda_{ijk}/(4\pi)^{\frac32} \, .
\ee
For the theory defined by \eqref{Lsix}, where
$\phi_i \to (\sigma,\vphi_i)$,   ${\hat \lambda}_{ijk}
{\hat \lambda}_{ijk} = 3N \, {\hat g}^2 + {\hat \lambda}{}^2$
and to lowest order $\beta\hskip 0.1pt{}_{\hat g} = - \half \vep \, {\hat g}
+ \tfrac{1}{12} (N-8)\,  {\hat g}^3 -  {\hat g}^2 {\hat \lambda}
+ \tfrac{1}{12} \, {\hat g}\,  {\hat \lambda}^2$ ,
$\beta\hskip 0.1pt \raisebox{-1.5 pt} {$\scriptstyle {\hat \lambda}$} =
-  \vep \, {\hat \lambda}  -  N \,  {\hat g}^3  + \tfrac14 N\,   {\hat g}^2 {\hat \lambda}
- \tfrac{3}{4} \, {\hat \lambda}^3$,
so that at the fixed point, to leading order in $\vep, 1/N$,
${\hat g}^2 = 6\vep /N, \, {\hat \lambda}{}^2 = 6^3\vep/N$ and hence
in \eqref{Cfree}, \eqref{Lsix} gives for the CFT at $d=6-\vep$ for large $N$
\be
C_{T,1} = 1 - \tfrac74 \, \vep \, ,
\ee
agreeing with the perturbative flat space calculation in \cite{Kleb4} and
also the expansion of the large $N$ result.  The corresponding results for
four-dimensional renormalisable theories were obtained some time ago. For
$n_A$ gauge fields, with a simple gauge group and coupling $g$, Dirac
fermions and Yukawa, scalar interactions ${\bar \psi} \hskip 0.5pt Y_i \psi
\, \phi_i$, $\tfrac{1}{24} \, \lambda_{ijkl} \phi_i \phi_j \phi_k \phi_l$
the results obtained in \cite{Hathrell1, Hathrell2, Freeman, Cappelli} by
expanding about flat space and, using heat kernel methods for a curved
background, \cite{Jack2, analogs} give\footnote{In terms of some previous
literature $c= - 16\pi^2 \beta_a$. In \cite{Jack2} in the final result a
misprint in (4.16) is corrected by taking $\frac19\to \frac29$.}
\be
c_1 = - \tfrac29 \big ( C - \tfrac78 \, R_\psi - \tfrac14 R_\phi \big ) n_A \, {\hat g}^2
- \tfrac{1}{24}\, \tr({{\hat Y}_i {\hat Y}_i} )
- \tfrac{1}{32\times 27} \, {\hat \lambda}_{ijkl} {\hat \lambda}_{ijkl} \, ,
\label{cone}
\ee
for  ${\hat g} = g/4\pi$,  ${\hat Y}_i = Y_i/4\pi$, ${\hat \lambda}_{ijkl}
= \lambda_{ijkl}/16\pi^2$ and we take the spinorial trace $\tr(
{\mathds{1)}} =4$. The conventions in \eqref{cone} for $C,R_\psi,R_\phi$
are such that the lowest order gauge $\beta$-function becomes $\beta_{\hat
g} = - \half \vep \, {\hat g} + \tfrac13( 11 C - 4 R_\psi - \half R_\phi)
{\hat g}^3 $.  For the $O(N)$ scalar theory, with interaction $\tfrac18
\,\lambda (\vphi^2)^2$ and $\beta\hskip 0.1pt \raisebox{-1.5 pt}
{$\scriptstyle {\hat \lambda}$} = - \vep \, {\hat\lambda} + (N+8) \,
{\hat\lambda}^2$, this was shown in \cite{Petkou2} to give a ${\rm
O}(\vep^2)$ contribution to $c$ at the RG fixed point in $4-\vep$
dimensions, $5!\,c_1 = - \tfrac{5}{12} N(N+2)/(N+8)^2 \, \vep^2$,
consistent with large $N$ results.  For the Gross--Neveu model starting
from \eqref{Lfour} the one-loop $\beta$-functions are
$\beta\hskip0.2pt{}_{\hat g} = - \half \vep \, {\hat g} + (2N+3) {\hat
g}^3$,  $\beta\hskip 0.1pt \raisebox{-1.5 pt} {$\scriptstyle {\hat
\lambda}$} = - \vep \, {\hat\lambda} + 3\, {\hat\lambda}^2 + 8N \, {\hat
\lambda}\, {\hat g}^2 - 48N \, {\hat g}^4$.  At the fixed point to leading
order for large $N$ from the Yukawa terms in \eqref{cone} $5! \, c_1 = -
5N/(4N+6) \, \vep$, which agrees with explicit calculations and the
expansion of the large $N$ Gross--Neveu result for $C_T$ in \cite{Kleb4}.

For scalar and fermion theories large $N$ methods allow non-trivial CFTs to
be formally defined for general dimensions $d$ which interpolate between
physical theories for $d$ an integer. The situation is less clear for gauge
theories since maintaining conformal invariance and gauge invariance is
more difficult, as was demonstrated in section 4. For a gauge-invariant
quantum field theory the energy-momentum tensor in general contains
contributions arising from the gauge fixing and ghost terms in the action.
However these are BRS exact and do not contribute to correlation functions
for gauge-invariant operators so that the calculations of $C_T$ in
\cite{Pet, Buchel} did not take account of them. In section 4 the gauge
fixing terms made a contribution to the energy-momentum tensor \eqref{EM0}
whose effect did not disappear in correlation functions when they
notionally decoupled. It is an open question whether the ghost
contributions to $T^{\mu\nu}$ could be extended to general $d$ while
maintaining conformal invariance. Their contributions to $C_T$ should
account for the difference between $C_T$ in \eqref{CTn} and the
corresponding gauge theory result when $d=2n$.

\bigskip
\noindent{\it Note Added}
\medskip

For higher-derivative scalar theories calculations of $C_T$ have also been
carried out by Guerrieri {\it et al.}~\cite{Guerr}, who further considered the
theory with eight derivatives. Prompted by their discussion there is a quick
derivation of $C_{T,\vphi,2p}$, agreeing with \eqref{Cscalar} for any $p$,
based on known results for conformal partial-wave expansions of four-point
functions as a sum over conformal primaries.  Using Mellin transform
methods Fitzpatrick and Kaplan~\cite{Fitzpatrick} obtained a conformal
partial-wave expansion for the four-point function of generalised free
fields in any dimension $d$.  Their results give the expansion
\be
u^\delta = \sum_{\ell,\tau=0}^\infty c_{\tau,\ell} \;
G_{2\delta+2\tau + \ell, \ell}(u,v) \, ,
\label{cpw}
\ee
where $u,v$ are the standard conformal invariants and
$G_{\Delta,\ell}(u,v)$ are the conformal partial waves for a conformal
primary operator with scaling dimension $\Delta$ and spin $\ell$ and its
descendants.  The partial waves are normalised here so that
$G_{\Delta,\ell}(u,v) \sim u^{\frac12(\Delta-\ell)} (-1+v)^\ell$ as $u\to
0, \, v\to 1$. From~\cite{Fitzpatrick},
\be
c_{\tau,\ell} = \frac{\big ( (\delta - \half d +1)_\tau \,
(\delta)_{\ell+\tau} \big )^2}{\ell!\, \tau! \, (\ell+\half d)_\tau \,
(2\delta+\tau-d+1)_\tau \, ( 2\delta + \ell+\tau - \half d)_\tau \,
(2\delta+2\tau + \ell -1)_\ell } \, .
\ee
The four-point function for free fields $\langle \vphi(x_1) \, \vphi(x_2)
\, \vphi(x_3) \, \vphi(x_4) \rangle $ defines a function of the conformal
invariants $F(u,v) = N_\vphi{\!}^2 \big ( 1 + u^{\Delta_\vphi} +
(u/v)^{\Delta_\vphi} \big ) $, where $N_\vphi$ is the coefficient for the
two-point function, which may be expanded in terms of conformal partial
waves using \eqref{cpw}, for $u\to u/v$ $c_{\tau,\ell} \to (-1)^\ell
c_{\tau,\ell}$. Coresponding to the operator product \eqref{opeI},
\be
\vphi(x) \, \vphi(0) \sim - \frac{N_\vphi}{C_{T,\vphi}} \,
\frac{d \, \Delta_\vphi} {d-1} \,
\frac{1}{(x^2)^{\Delta_\vphi - \frac12 d +1 } } \,
{x}_\mu {x}_\nu \, T_\vphi^{\mu\nu}(0) \, .
\ee
Since ${x}_{12\,\mu}\hspace{0.8pt} {x}_{34\, \nu} \, I^{\mu\nu}(x_{24})
\big /(x_{12}{\!}^2 x_{34}{\!}^2)^\frac12 \sim
-(1-v)/(2\sqrt u)$ as $x_{12}, x_{34}\to 0$, we must have
\be
c_{p-1,2} \Big |_{\delta = \frac12(d-2p)} = \bigg (
\frac{d \, \Delta_{2p}}{d-1} \bigg )^2 \, \frac{1}{8 \, C_{T,\vphi,2p} }
= (-1)^{p-1} \, \frac{d (d-2p)^2}{32 \, p (d-1)} \,
\frac{(\frac12 d-p+1)_{p-1}} {(\half d+2)_{p-1}} \, .
\ee
This gives a result for $C_{T,\vphi,2p}$ identical to that in
\eqref{Cscalar}.  The restriction on $\delta$ is of course necessary to
ensure that the expansion contains a conformal primary contribution which
may be identified with the energy-momentum tensor, for this case in
\eqref{cpw} $\tau \le p-1$ and the operators which contribute have
$\Delta-\ell = d+ 2(\tau -p)$ and are expressible as $\vphi (
{\tikzoverset{\text{\tiny$\leftrightarrow$}}{\pr}})^\ell \, (\pr^2)^\tau
\vphi$.

These results can easily be extended to consider the differing conserved
currents present in these free theories. If we allow $\vphi \to \vphi_i$,
there is a conserved current $J_{\vphi,ij}^\mu = - J_{\vphi,ji}^\mu$ with
\be
\big \langle J_{\vphi,ij}^{\mu}(x) \, J_{\vphi,kl}^{\nu}(0)
\big \rangle
= C_{J,\vphi} \,
\frac{1}{(x^2)^{d-1}} \, I^{\mu\nu}(x) \, ( \delta_{ik} \delta_{jl}
- \delta_{il} \delta_{jk} ) \, .
\ee
For its contribution in the operator product expansion we have
\be
\vphi_i(x) \, \vphi_j(0) \sim - \frac{N_\vphi}{C_{J,\vphi}} \,
\frac{1}{(x^2)^{\Delta_\vphi - \frac12 d +1 } } \,
{x}_\mu  \, J_{\vphi,ij}^{\mu}(0) \, .
\ee
The four-point function now has the form
$F_{ijkl}(u,v) = N_\vphi{\!}^2 \big ( \delta_{ij} \delta_{kl} +
\delta_{ik} \delta_{jl} \, u^{\Delta_\vphi} +
\delta_{il} \delta_{jk} \, (u/v)^{\Delta_\vphi} \big ) $, and using
\eqref{cpw} we find
\be
\frac{1}{C_{J,\vphi,2p}} = 2\,c_{p-1,1} \Big |_{\delta = \frac12(d-2p)}
= (-1)^{p-1} \, \frac{1}{p} \, \frac{(\frac12 d-p)_p}
{ (\frac12 d+1)_{p-1} } \, .
\ee

\ack{For some of our computations we have used \emph{Mathematica} with the
package \texttt{FeynCalc}~\cite{Mertig, Shtabovenko}. AS would like to
thank Brian Henning, David Poland, and Siddharth Prabhu for useful
discussions. The research of AS is supported in part by the National
Science Foundation under Grant No.~1350180.}

\bibliography{nonUnit}

\end{document}